\documentclass[aps,prb,twocolumn,amsmath,amssymb,superscriptaddress]{revtex4}
\usepackage{graphicx}
\usepackage{xcolor}
\colorlet{RED}{red}
\usepackage{amssymb}
\usepackage{natbib}
\usepackage{calc}
\usepackage{SIunits}
\usepackage{textcomp}
\graphicspath{ {./Figures/} }

\begin{document}
\title{Spin order and excitations in the collinear kagome antiferromagnet Y$_{0.5}$Fe$_3$Sn$_3$}

\author{Jin Ding}
\thanks{These authors made equal contributions to this paper.}
\affiliation{Beijing National Laboratory for Condensed Matter Physics, Institute of Physics, Chinese Academy of Sciences, Beijing 100190, China}
\affiliation{University of Chinese Academy of Sciences, Beijing 100049, China}
\author{Yang Liu}
\thanks{These authors made equal contributions to this paper.}
\affiliation{Beijing National Laboratory for Condensed Matter Physics, Institute of Physics, Chinese Academy of Sciences, Beijing 100190, China}
\affiliation{University of Chinese Academy of Sciences, Beijing 100049, China}
\author{Feihao Pan}
\affiliation{J\"{u}lich Centre for Neutron Science at Heinz Maier-Leibnitz Zentrum, Forschungszentrum J\"{u}lich GmbH, Lichtenbergstrasse 1, D-85747 Garching, Germany}
\author{Yinghao Zhu}
\affiliation{J\"{u}lich Centre for Neutron Science at Heinz Maier-Leibnitz Zentrum, Forschungszentrum J\"{u}lich GmbH, Lichtenbergstrasse 1, D-85747 Garching, Germany}
\author{Yixi Su}
\affiliation{J\"{u}lich Centre for Neutron Science at Heinz Maier-Leibnitz Zentrum, Forschungszentrum J\"{u}lich GmbH, Lichtenbergstrasse 1, D-85747 Garching, Germany}
\author{Stanislav E. Nikitin}
\affiliation{PSI Center for Neutron and Muon Sciences, 5232 Villigen PSI, Switzerland}
\author{Ursula B. Hansen}
\affiliation{Institut Laue-Langevin, 71 Avenue des Martyrs, 38042 Grenoble, France}
\author{Mechthild Enderle}
\affiliation{Institut Laue-Langevin, 71 Avenue des Martyrs, 38042 Grenoble, France}
\author{Zhilun Lu}
\affiliation{School of Chemical and Process Engineering, University of Leeds, Leeds, West Yorkshire LS2 9JT, England, United Kingdom}
\author{Linxuan Song}
\affiliation{Beijing National Laboratory for Condensed Matter Physics, Institute of Physics, Chinese Academy of Sciences, Beijing 100190, China}
\author{Yong-Chang Lau}
\affiliation{Beijing National Laboratory for Condensed Matter Physics, Institute of Physics, Chinese Academy of Sciences, Beijing 100190, China}
\affiliation{University of Chinese Academy of Sciences, Beijing 100049, China}
\author{Xiaoyan Ma}
\affiliation{Beijing National Laboratory for Condensed Matter Physics, Institute of Physics, Chinese Academy of Sciences, Beijing 100190, China}
\affiliation{University of Chinese Academy of Sciences, Beijing 100049, China}
\author{Enke Liu}
\email{ekliu@iphy.ac.cn}
\affiliation{Beijing National Laboratory for Condensed Matter Physics, Institute of Physics, Chinese Academy of Sciences, Beijing 100190, China}
\author{Huiqian Luo}
\email{hqluo@iphy.ac.cn}
\affiliation{Beijing National Laboratory for Condensed Matter Physics, Institute of Physics, Chinese Academy of Sciences, Beijing 100190, China}

\date{\today}
%\pacs{75.50.Ee, 67.57.Lm, 75.30.Gw, 25.40.Fq, 25.40.Dn}

\begin{abstract}
Kagome lattices host rich magnetic, topological, and transport phenomena owing to their geometric frustration and strong spin correlations. Here we present a systematic neutron scattering study of the kagome metal Y$_{0.5}$Fe$_3$Sn$_3$, which orders into a collinear $c$-axis antiferromagnetic (AF) state below $T_N =$ 551 K. An additional anomalous Hall effect (AHE) contribution beyond the conventional magnetization-scaling response is observed well below $T_N$ under an in-plane magnetic field $\mathbf{B} \parallel ab$. Elastic neutron scattering demonstrates that the Fe moments preserve their collinear $c$-axis AF alignment down to 1.5 K at zero field. Inelastic neutron scattering reveals a distinctive Y-shape spin excitation dispersion exhibiting a two-gap structure below 10 meV, likely arising from two nondegenerate spin-wave branches. Polarized neutron measurements further show that the low-temperature spin excitations are strongly anisotropic: the in-plane component $M_{ab}$ dominates while the $c$-axis component $M_c$ is nearly absent. Upon warming, $M_c$ gradually grows markedly but remains much weaker than $M_{ab}$ even at 300 K. We propose that the unconventional AHE originates from a nonzero scalar spin chirality induced by field-driven spin canting either in the static or dynamic configurations due to gapless excitations. These findings establish the collinear AF kagome metal Y$_{0.5}$Fe$_3$Sn$_3$ as a platform for chirality-driven AHE without spontaneous spin reorientation.

\end{abstract}

\maketitle
\section{Introduction}
Kagome metals are ideal hosts for intriguing electronic band structures including flat bands, Van Hove singularities, and Dirac points \cite{BBernevig2022,Yin2022,YWang2023,Chowdhury2023,QWang2024,MLi2025}. The spin interactions in the kagome lattice further drive many emergent quantum phenomena, such as anomalous Hall effect (AHE), spin chirality, quantum spin liquid, magnon topology, and unconventional superconductivity \cite{Kuroda2017,Ye2018,ZYLin2018,EKLiu2018,CLiu2021,Onose2010,Balents2010,Chisnell2015,ZZeng2024,LHHu2025,SGuo2025,KJiang2023,YLiu2024,YZLi2026}. The AHE is usually interpreted in terms of two distinct mechanisms: an extrinsic contribution induced by impurity scattering in the presence of spin-orbit coupling, and an intrinsic contribution arising from the Berry curvature of Bloch electrons in momentum space. Substantial progress has been made in disentangling the AHE across diverse kagome platforms \cite{EHHall1881,EMPugh1932,Nagaosa2010,HWeng2015,Bandyopadhyay2024}. While both intrinsic and extrinsic AHE have been observed in many ferromagnetic (FM) or ferrimagnetic kagome compounds such as Co$_3$Sn$_2$S$_2$, Fe$_3$Sn$_2$, and $R$Mn$_6$Sn$_6$ ($R$=rare-earth elements) \cite{Kuroda2017,Ye2018,ZYLin2018,EKLiu2018,gventurini1996,bmalaman1999,Ghimire2020,gdhakal2021,llgao2021,kfruhlingt2024,Riberolles2024b,yzhou2025}, intrinsic AHE has been revealed in noncollinear antiferromagnets such as Mn$_3$Sn and Mn$_3$Ge, where time-reversal symmetry is broken despite nearly compensated sublattice moments\cite{aknayak2016,nkiyohara2016,tsuzuki2016,nhsung2018,xkli2023,ppark2018,csurgers2017}. Recently, CsCr$_3$Sb$_5$ is also proposed to host an AHE in its altermagnetic state with even-parity spin splitting \cite{ccxu2025}. The exploration of AHE in systems with nearly vanishing net magnetization is of particular interest owing to their potential advantages for spintronic applications.

Besides the conventional intrinsic AHE with $\rho^A_{yx}\propto M$ driven by momentum-space Berry curvature—typically in ferromagnetic systems—a Berry curvature generated in real space can give rise to another type of Hall response, often termed as the topological Hall effect \cite{Smejkal2022,Metalidis2006,Neubauer2009,Ueland2012,Bruno2004,Ueda2012,Nagaosa2013,Nakatsuji2015,Kurumaji2019,Fujishiro2021,JShen2023,Abe2024,JJMa2025,RXiao2025}. Specifically, when charge carriers traverse noncoplanar or noncollinear spin textures such as skyrmions or canted magnetic structures, the local scalar spin chirality (SSC) $\chi=\mathbf{S}_i \cdot (\mathbf{S}_j \times \mathbf{S}_k)$ (where $\mathbf{S}_i,\mathbf{S}_j$, and $\mathbf{S}_k$ are three neighboring spins) acts as an emergent magnetic field and generates a Hall signal beyond the conventional magnetization-scaling paradigm \cite{Yang2011,Chen2014,Denisov2018,Ishizuka2021,Kolincio2023,Roychowdhury2024}. Moreover, thermally enhanced or field-tuned spin fluctuations can contribute to the AHE even in the spin-disordered state, provided the time-averaged SSC $\langle\mathbf{S}_i \cdot (\mathbf{S}_j \times \mathbf{S}_k)\rangle$ is nonzero—as demonstrated in Nd$_3$Ru$_4$Al$_{12}$, Gd$_3$Ru$_4$Al$_{12}$, HoAgGe, etc. \cite{Hou2017,Nagaosa2018,Kato2019,Kolincio2021}. A rare realization of unconventional AHE (UAHE) in a collinear antiferromagnetic (AF) state has recently been observed in YbFe$_6$Ge$_6$, where the AHE emerges in the spin-reoriented phase (from the $c$ axis to the kagome plane) in conjunction with gapless spin excitations far below the AF transition at $T_N \approx$ 500 K \cite{Yao2025,TMazet2000a,Avila2005,DHRyan2010}.

Here we focus on another collinear kagome antiferromagnet Y$_{0.5}$Fe$_3$Sn$_3$ (Fig. 1) \cite{Liu2023,Idrissi1991,XLRao1998,Cadogan2000,TMazet2000b,TMazet2001}, a cousin compound of YbFe$_6$Ge$_6$. Previous reports on the 166-type kagome family of $R$$T$$_6$Sn$_6$ ($R$ = rare-earth elements, $T$ = transition metal elements) have revealed rich magnetic states due to the competing interactions between $T$ kagome and $R$ triangular layers, such as spin-helix order, band-crossing spin waves, and topological Dirac magnons \cite{Liu2023,Idrissi1991,XLRao1998,Cadogan2000,TMazet2000b,TMazet2001,HZhang2020,QWang2021,RDally2021,WMa2021,HZhang2022,YLee2023,JYin2020,Riberolles2022,Riberolles2024,Mielke2022,Riberolles2023,Jones2024,ZHuang2024}. Both ferrimagnetic ($R$ = Gd, Tb, Dy, Ho) and AF ($R$ = Sc, Y, Er, Tm, Yb, Lu) ground states are observed in the $R$Mn$_6$Sn$_6$ family \cite{WMa2021,HZhang2022,YLee2023,JYin2020,Riberolles2022,Riberolles2024,Mielke2022,Riberolles2023,Jones2024,ZHuang2024}, and some of these magnetic phases experience a spin-reorientation transition on cooling away from the critical temperature \cite{JYin2020,Riberolles2022,Riberolles2024,Mielke2022,Riberolles2023,Jones2024,ZHuang2024}. Thus the spins can align either within the kagome plane or perpendicular to it [Figs. 1(a) and (b)]. In principle, the collinear order (whether FM or AF) necessarily yields zero static SSC, whereas a noncollinear order may produce a nonzero SSC under the broken combined space-inversion and time-reversal ($\mathcal{IT}$) symmetry.
The Fe-based YFe$_6$Sn$_6$ crystallizes in a YCo$_6$Ge$_6$-type structure with Fe-kagome layers (space group $P6/mmm$) \cite{Liu2023,Idrissi1991,XLRao1998,Cadogan2000,TMazet2000b,TMazet2001}. Its structure can be viewed as two interpenetrating HfFe$_6$Ge$_6$-type sublattices, one shifted by $c/2$ relative to the other, with Y and Sn sites at only half occupancy \cite{Weiland2020}. Consequently, the unit cell is halved compared with the canonical 166-type kagome compounds, yielding the stoichiometric formula Y$_{0.5}$Fe$_3$Sn$_3$ \cite{Liu2023}. Magnetic susceptibility measurements indicate an AF order with $T_N =$ 551 K, where the ordered moments likely point either along the $c$ axis or within the kagome plane [Figs. 1(c) and (d)]. Previous electronic transport measurements revealed a single-carrier AHE exclusively under an in-plane magnetic field ($\mathbf{B} \parallel ab$) \cite{Liu2023}. Through a detailed analysis of the field-dependent anomalous Hall resistivity combined with magnetization data, we further identify an additional AHE contribution ($\Delta \rho^A_{yx}$) that does not follow the net magnetization. This unconventional AHE closely resembles the case in YbFe$_6$Ge$_6$, albeit with distinct temperature ($\approx$ 150 K) and field ($\approx$ 2 T) thresholds.

Our elastic neutron scattering experiments on Y$_{0.5}$Fe$_3$Sn$_3$ confirm its collinear AF structure, where the magnetic moments keep the alignment along the $c$ axis down to $T=$ 1.5 K without any reorientation, in contrast to other 166-type materials \cite{Yao2025,TMazet2000a,Avila2005,DHRyan2010,Liu2023,Idrissi1991,XLRao1998,Cadogan2000,TMazet2000b,TMazet2001,HZhang2020,QWang2021,RDally2021,WMa2021,HZhang2022,YLee2023}. Inelastic neutron scattering (INS) measurements reveal unique Y-shape dispersions in the low-energy spin waves, and the intensity below 10 meV abnormally increases upon warming up to 300 K. Using polarized analyses of INS, we demonstrate that the spin excitations are highly anisotropic in spin space consisting of dominant in-plane component $M_{ab}$ and nearly zero $c$-axis component $M_{c}$ above 3 meV at 1.5 K, and $M_{c}$ gradually evolves to $E=$ 14 meV at $T=$ 300 K. These observations point to anisotropic spin interactions as a viable route to SSC-driven AHE in the collinear kagome antiferromagnet.

\begin{figure}[tbp]
	
	\includegraphics[width=0.45\textwidth]{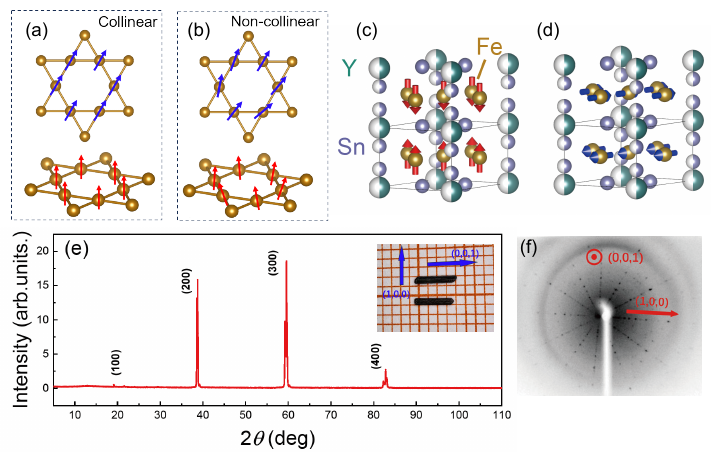}
	\caption{(a), (b) Collinear and noncollinear magnetic orders in a kagome lattice with in-plane and out-of-plane spin orientations. 
			(c), (d) Crystal structure and possible antiferromagnetic orders of Y$_{0.5}$Fe$_3$Sn$_3$ with spin orientation either along the $c$ axis or in the $ab$ plane \cite{Liu2023}. Here Y$_{0.5}$Fe$_3$Sn$_3$ shows the YCo$_6$Ge$_6$-type structure composed of two HfFe$_6$Ge$_6$-type subunits, with one offset by $c/2$ relative to the other and half occupancy of Y and Sn atoms.
			(e) Single-crystal x-ray diffraction pattern showing $(H, 0, 0)$ reflections. Inset: Typical one-dimensional-like crystals sized about 1 $\times$ 4 $\times$ 0.5 mm $^3$.
			(f) X-ray Laue reflection pattern of a typical crystal at room temperature. The crystallographic axes of $(1, 0, 0)$ and $(0, 0, 1)$ are marked by red arrows.
	}
\end{figure}

\section{Experiments}

\subsection{Crystal growth and characterization}
High-quality single crystals of Y$_{0.5}$Fe$_3$Sn$_3$ were grown by a Sn self-flux method following the previously reported procedure \cite{Liu2023}. High-purity Y ingots (99.9\%), Fe pieces (99.99\%), and Sn pellets (99.999\%) were mixed in a molar ratio of 1:6:10, placed in an alumina crucible under an argon atmosphere, and subsequently sealed under vacuum in a quartz tube. The mixture was heated to 1100 $^{\circ}$C, held at this temperature for 24 h to ensure homogenization, slowly cooled to 700 $^{\circ}$C, and then centrifuged to remove the excess Sn flux. The obtained crystals typically show a hexagonal thin-pillar morphology with the long axis along the crystallographic $c$ direction. The crystalline quality was checked by single-crystal x-ray diffraction (XRD) and Laue diffraction, as shown in Figs. 1(e) and 1(f). The XRD pattern in Fig. 1(e) displays only sharp $(H, 0, 0)$ reflections, and the sixfold-symmetric Laue pattern in Fig. 1(f) confirms the high crystalline quality and well-defined orientation of our crystals.

\subsection{Transport and magnetization measurements}
Electrical transport measurements were carried out using a physical property measurement system. The longitudinal resistivity $\rho_{xx}$ was measured with current applied along the $c$ axis, and Hall resistivity $\rho_{yx}$ was measured with magnetic field oriented in the kagome plane ($\mathbf{B}\parallel ab$). The anomalous Hall conductivity was extracted from the field-dependent Hall response together with the zero-field longitudinal resistivity, following the analysis detailed in the Supplemental Material \cite{supplementary}. Magnetization measurements were performed using a magnetic property measurement system and a high-temperature vibrating sample magnetometer up to 800 K, with a field of $B=0.1$ T applied along both the $ab$ and $c$ directions.

\subsection{Neutron scattering experiments}
For neutron scattering measurements, about 4.6 g of crystals was co-aligned in the $[H, 0, L]$ scattering plane using hydrogen-free glue on aluminum plates. The co-alignment quality was verified by rocking scans of nuclear Bragg peaks, with full-width-at-half-maximum below $2^\degree$ for both in-plane and out-of-plane directions \cite{supplementary,Collins1967,hqluo2012,xylu2013,dgong2018,wlzhang2019}. A YCo$_6$Ge$_6$-type hexagonal unit cell was adopted, with lattice parameters $a=b=5.373$ \AA, $c=4.453$ \AA, $\alpha=\beta=90\degree$, and $\gamma=120\degree$, where $c$ is half of that in typical 166-type compounds. The momentum transfer was denoted as $\mathbf{Q}=H\mathbf{a}^*+K\mathbf{b}^*+L\mathbf{c}^*$ in reciprocal lattice units, where $a^*=b^*=4\pi/\sqrt{3}a$ and $c^*=2\pi/c$ in the hexagonal lattice.

Unpolarized elastic and inelastic neutron scattering measurements were carried out on the thermal triple-axis spectrometer EIGER at the Paul Scherrer Institut, Switzerland \cite{UStuhr2017}. A fixed final neutron energy of $E_f=14.7$ meV ($k_f=2.662$ \AA$^{-1}$) was used in the energy-loss mode, together with a pyrolytic graphite filter and a double-focusing monochromator. Constant-energy scans ($Q$ scans) were analyzed after background subtraction and Bose-factor correction to obtain the imaginary part of the dynamic susceptibility $\chi^{\prime\prime}(\mathbf{Q},E)$. The extracted peak centers were then used to construct the spin-wave dispersion at low energy as shown in the Supplemental Material \cite{supplementary}.

\begin{figure}[tbp]
	
	\includegraphics[width=0.45\textwidth]{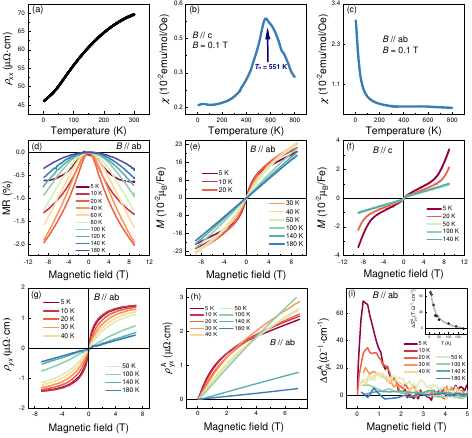}
	\caption{(a) Temperature dependence of the zero-field $c$-axis resistivity $\rho_{xx}$.
		(b), (c) Temperature dependence of the magnetic susceptibility for $\mathbf{B} \parallel c$ and $\mathbf{B} \parallel ab$ ($B=0.1$ T), respectively. The arrow marks the AF transition at $T_N = 551$ K.
		(d) Field dependence of the magnetoresistance with current $I \parallel c$ and magnetic field $\mathbf{B} \parallel ab$.
		(e), (f) Field dependence of the magnetization at different temperatures with magnetic field $\mathbf{B} \parallel ab$ and $\mathbf{B} \parallel c$, respectively.
		(g) Field dependence of total Hall resistivity $\rho_{yx}$ with $I \parallel c$ and $\mathbf{B} \parallel ab$ from $T=5$ K to 180 K.
		(h) Field dependence of conventional anomalous Hall resistivity due to net magnetization ($\rho^A_{yx}\propto M$) from $T=5$ K to 180 K.
		(i) Field dependence of additional AHC contributions $\Delta\sigma^A_{yx}$ from $T=5$ K to 180 K. Inset: Integrated intensity $\Delta S^A_{yx}$ vs temperature. 
	}
\end{figure}

Fully polarized neutron scattering measurements were performed on the IN20 thermal triple-axis spectrometer at the Institut Laue-Langevin, France, using the CryoPAD system together with the Heusler monochromator and analyzer \cite{Berna2005,IN20Exp2025}.  This spectrometer has a velocity selector to suppress higher-order wavelengths from the monochromator enabling increased flexibility and improved signal-to-noise ratio. The neutron polarization directions $x$, $y$, and $z$ were defined with $x\parallel\mathbf{Q}$ and $y\perp\mathbf{Q}$ within the scattering plane, and $z$ perpendicular to the scattering plane. By measuring the $xx$, $yy$, and $zz$ polarization channels in both spin-flip (SF) and non-spin-flip (NSF) configurations, the magnetic components $M_y$ and $M_z$ were separated from the coherent nuclear contribution $\sigma_{\mathrm{Nuc}}$ and the incoherent background $\sigma_{\mathrm{Inc}}$ \cite{Lipscombe2010,NQureshi2012,HLuo2013,CWang2013,YSong2017,CLiu2022}. Given the scattering geometry, $M_y=M_{\parallel}\sin^2\theta+M_c\cos^2\theta$ and $M_z=M_{\perp}$, where $\theta$ is the angle between the $y$ and $c$ directions, $M_c$ is the $c$-axis component, and $M_{\parallel}$ and $M_{\perp}$ are the two in-plane spin components alternatively denoted as $M_{ab}$ when they are equal. All raw polarized INS data and the analysis procedure are presented in the Supplemental Material \cite{supplementary}.
\begin{figure}[tbp]
	
	\includegraphics[width=0.45\textwidth]{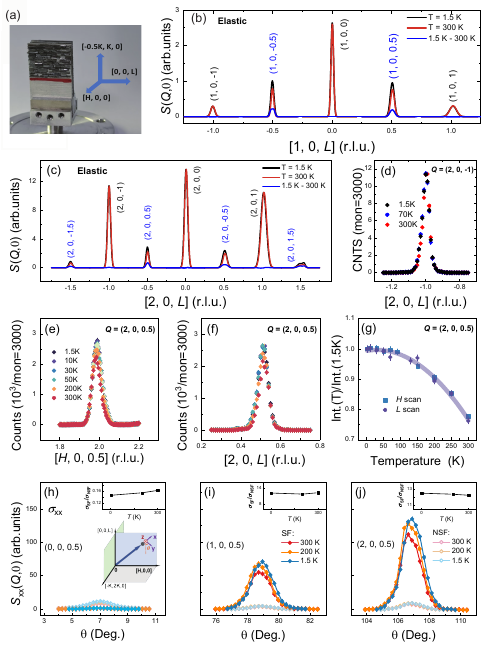}
	\caption{ (a) Photograph of the co-aligned single crystals used in our neutron scattering experiments. Blue arrows denote the crystallographic orientations in reciprocal space.
		(b), (c) Elastic neutron scattering results measured at EIGER for long $Q$ scans along the [1, 0, $L$] and [2, 0, $L$] directions. The blue curves are the difference of peak intensity between 1.5 and 300 K.
		(d) $L$ scans of the nuclear Bragg peak at $Q$ = (2, 0, -1) for $T=$ 1.5, 70, and 300 K measured at EIGER.
		(e), (f) Temperature dependence of elastic peak intensity at $Q$ = (2, 0, 0.5) for [$H$, 0, 0.5] and [2, 0, $L$] directions measured at EIGER.
		(g) Temperature dependence of integrated peak intensity both for $L$ scan and $H$ scan at $Q=$ (2, 0, 0.5).
		(h)-(j) SF and NSF intensities of elastic rocking curves measured at IN20 in the $xx$ polarization channel at $Q$ = (0, 0, 0.5), (1, 0, 0.5) and (2, 0, 0.5). Insets: Ratio of peak intensity between SF and NSF signals. The inset of panel (h) also shows the definition of spin-polarization directions ($x$, $y$, $z$) in reciprocal space at the measured momentum transfer $\mathbf{Q}$, where $\theta$ is the angle between $y$ and $L$ directions.
	}
\end{figure}

\section{Results}

\subsection{Electrical transport and magnetization}

We begin with the electrical transport and magnetization results presented in Fig. 2. The zero-field $\rho_{xx}(T)$ in Fig. 2(a) exhibits a metallic temperature dependence from 300 K down to 2 K, decreasing smoothly on cooling without any pronounced anomaly. The magnetic susceptibility is strongly anisotropic [Figs. 2(b) and 2(c)]: for $\mathbf{B} \parallel c$, $\chi_c(T)$ displays a clear cusp at $T_N \approx 551$ K, consistent with AF order with moments aligned predominantly along the $c$ axis, whereas for $\mathbf{B} \parallel ab$, $\chi_{ab}(T)$ evolves smoothly and becomes substantially larger at low temperatures. This is consistent with the absence of a low-temperature spin-reorientation signature in the bulk magnetization, since a spin reorientation while preserving the AF correlations would produce a pronounced drop in $\chi_{ab}$ and a step-like increase in $\chi_{c}$, as observed in YbFe$_6$Ge$_6$ \cite{Yao2025,TMazet2000a,Avila2005,DHRyan2010}. The in-plane magnetoresistance in Fig. 2(d), measured with $I \parallel c$ and $\mathbf{B}\parallel ab$, is negative over the entire measured field range and becomes more pronounced on cooling, pointing to an enhanced coupling between charge transport and the field-tuned magnetic state at low temperatures. The field-dependent magnetization for $\mathbf{B} \parallel ab$ tends to saturate at high fields in the low-temperature regime and becomes linear at elevated temperatures [Fig. 2(e)]. When the magnetic field is applied along the $c$ axis, the magnetization does not saturate owing to the anti-parallel arrangement of magnetic moments [Fig. 2(f)].
	
The Hall resistivity $\rho_{yx}$ for $\mathbf{B}\parallel ab$, shown in Fig. 2(g), is markedly nonlinear below 140 K and approaches saturation at high fields. Theoretically, $\rho_{yx}$ comprises three contributions:  $\rho_{yx}=\rho^O_{yx}+\rho^A_{yx}+\Delta\rho^A_{yx}$, where $\rho^O_{yx}$ is the ordinary Hall resistivity linear in the applied magnetic field $B$, $\rho^A_{yx}$ is the conventional anomalous Hall resistivity arising from the net magnetization ($\rho^A_{yx}\propto M$), and $\Delta\rho^A_{yx}$ represents the additional (unconventional) AHE contribution not captured by the first two terms \cite{Yao2025,wwang2022}. The corresponding additional anomalous Hall conductivity (AHC) $\Delta\sigma^A_{yx}$ follows from $\Delta\sigma^A_{yx}=\Delta\rho^A_{yx}/(\rho^2_{xx}+\rho^2_{yx})$. The conventional anomalous Hall resistivity $\rho^A_{yx}$ and the unconventional AHC $\Delta\sigma^A_{yx}$ are plotted in Figs. 2(h) and 2(i). It is of particular interest that Y$_{0.5}$Fe$_3$Sn$_3$ exhibits an unconventional AHE with a nonzero $\Delta\sigma^A_{yx} \approx$ 20 $\Omega^{-1}$cm$^{-1}$ at $T=$ 5 K, closely resembling the behavior reported in YbFe$_6$Ge$_6$ \cite{Yao2025}. The principal differences lie in the temperature ($\approx$ 150 K) and field ($\approx$ 2 T) thresholds.

\subsection{Elastic neutron scattering}

To confirm the AF structure of Y$_{0.5}$Fe$_3$Sn$_3$, elastic neutron scattering measurements were carried out on the co-aligned crystal assembly [Fig. 3(a)].  As shown in Figs. 3(b)-3(g), the elastic $Q$ scans along $[1, 0, L]$ and $[2, 0, L]$ unambiguously reveal magnetic Bragg peaks at $Q=(1, 0, \pm 0.5)$, $(2, 0, \pm 0.5)$, and $(2, 0, \pm 1.5)$, whose intensity decreases progressively upon warming from 1.5 to 300 K. By contrast, all measured nuclear Bragg peaks at $Q=(1, 0, 0)$, $(1, 0, \pm 1)$, $(2, 0, 0)$, and $(2, 0, \pm 1)$ show no discernible temperature dependence. These observations are fully consistent with a collinear AF structure characterized by a propagation vector $k_m=$ (0, 0, 0.5). Owing to the temperature limit of the Orange cryostat, we could not reach the N\'{e}el temperature $T_N=$ 551 K. Nonetheless, the peak intensity at $Q=(2, 0, 0.5)$ follows the expected behavior of a magnetic order parameter [Figs. 3(e)-3(g)].

The $c$-axis AF order can be rigorously tested by polarized elastic neutron scattering, which separates the magnetic and nuclear scattering into the SF and NSF channels, respectively \cite{supplementary}. Figures 3(h)-3(j)  display the cross section $\sigma_{xx}$ in both the SF and NSF channels at $Q_1=$ (0, 0, 0.5), $Q_2=$ (1, 0, 0.5), and $Q_3=$ (2, 0, 0.5) for $T=$ 1.5, 200, and 300 K. Dominant SF signals are observed at $Q_2$ and $Q_3$, confirming the magnetic origin of these peaks. At $Q_1=$ (0, 0, 0.5), where $\theta=90^{\degree}$, $\sigma^{\mathrm{SF}}_{xx}(Q_1, 0)$ mainly probes the in-plane static moment components $M_{\parallel}(Q_1, 0)$ and $M_{\perp}(Q_1, 0)$, both of which are essentially zero within experimental uncertainty. A weak nuclear scattering signal is also detected in $\sigma^{\mathrm{NSF}}_{xx}(Q_1, 0)$, since $Q_1$ actually corresponds to the (0, 0, 1) nuclear peak in the 166-type structure. The similar ratios $\sigma^{\mathrm{SF}}_{xx}/\sigma^{\mathrm{NSF}}_{xx}$ observed from 1.5 to 300 K demonstrate that the Fe moments remain fully aligned along the $c$ axis without any spin reorientation throughout this temperature range. By comparing the integrated intensities of the SF and NSF channels to separate nuclear and magnetic scattering, the static ordered moment at 1.5 K is estimated to be about 0.34 $\mu_B/$Fe \cite{supplementary}.  It should be emphasized that the above analysis is strictly valid only at zero magnetic field. Under an in-plane field $\mathbf{B} \parallel ab$, a weak ferromagnetic component may be induced by tilting of the Fe moments toward the kagome plane, if the single-ion spin anisotropy is sufficiently small. Such a field-induced FM component, however, would contribute solely to the conventional AHE term $\rho^A_{yx}$.

\begin{figure}[tbp]
	\includegraphics[width=0.42\textwidth]{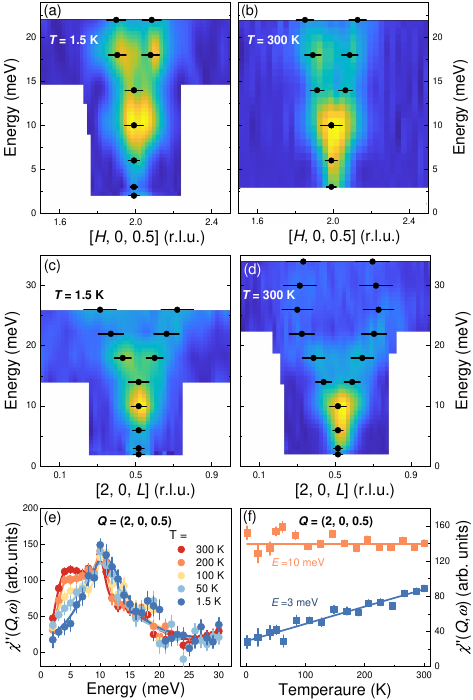}
	\caption{Unpolarized INS results collected at EIGER. 
		(a)-(d) Dispersions along the $H$ and $L$ directions for $Q=$ (2, 0, 0.5) at $T = 1.5$ and 300 K. The color gradient (blue to yellow) represents the intensity distributions. The black dots mark the peak positions obtained from Gaussian fittings \cite{supplementary}.
		(e) Energy scans at $Q$= (2, 0, 0.5) at $T=$ 1.5, 50, 100, 200, and 300 K. Here the intensity is converted to the local dynamic susceptibility $\chi^{\prime\prime}(Q, E)$ after subtracting the backgrounds at $Q$= (2, 0, 0) and (2, 0, 1) and further corrected by the Bose factor \cite{supplementary}. The solid lines are guides to the eye. 
		(f) Temperature dependence of $\chi^{\prime\prime}(Q, E)$ at $E=$ 3 and 10 meV. 
	}
\end{figure}

\subsection{Unpolarized inelastic neutron scattering}

Having established the collinear $c$-axis AF order with static moments and the absence of spin reorientation at low temperature, we now turn to the spin-wave excitations probed by unpolarized INS at EIGER \cite{UStuhr2017}. By performing $Q$ scans along both the $H$ and $L$ directions around the AF Bragg peak $Q=$ (2, 0, 0.5) at selected energies, we map out the low-energy spin-wave dispersions [Figs. 4(a)-(d)]. Remarkably, all dispersions (especially along the $L$ direction) exhibit a Y-shape profile with a crossing point at about $E=$ 10 meV, in striking contrast to the V-shape dispersion expected from conventional linear spin-wave theory. This feature likely arises from the superposition of two nondegenerate spin-wave branches with unequal spin gaps at low energy, a scenario that calls for further clarification by high-resolution INS experiments. We also note that the low-energy excitation intensity is counterintuitively stronger at 300 K than at 1.5 K. To investigate this further, we performed energy scans at several temperatures and extracted the local dynamic susceptibility $\chi^{\prime\prime} (Q, E)$. As shown in Fig. 4(e), the energy scans display a $\lambda$-shape profile that naturally divides the data into two regimes: upon warming to 300 K, the high-energy excitations remain nearly temperature independent, whereas the low-energy excitations below 10 meV grow anomalously and tend to develop a distinct mode near 4 meV. If the low-energy spectral weight indeed shifts to around 4 meV, a spin gap likely opens below 2 meV (beyond the resolution limit of EIGER) at $T=$ 300 K [Fig. 4(e)], reminiscent of the behavior reported in YbFe$_6$Ge$_6$ \cite{Yao2025}. Indeed, the low-energy $\chi^{\prime\prime} (Q, E)$ at 3 meV increases anomalously with temperature as the gap opens, whereas the overlapping signals at 10 meV remain nearly temperature independent [Fig. 4(f)]. The neutron scattering cross section $S (Q, E)$ does not appear to follow the Bose population factor at either energy, further underscoring the unconventional nature of these spin excitations \cite{supplementary}.

\subsection{Polarized inelastic neutron scattering analyses}

To further elucidate this anomalous temperature dependence, we employed polarized neutron scattering on IN20 to probe the spin anisotropy of the low-energy excitations \cite{Berna2005,IN20Exp2025}. The polarized results are presented in Figs. 5 and 6, and a detailed account of the analysis procedure is provided in the Supplementary Material \cite{supplementary}.  Since all three scattering cross sections $\sigma_{xx}$, $\sigma_{yy}$, and $\sigma_{zz}$ were measured in both SF and NSF channels, the clean magnetic excitation components $M_y$ and $M_z$ can be isolated by subtracting the nuclear scattering $\sigma_{\mathrm{Nuc}}$ and the incoherent background $\sigma_{\mathrm{Inc}}$ \cite{Lipscombe2010,NQureshi2012,HLuo2013,CWang2013,YSong2017,CLiu2022}. Energy scans were performed at three wave vectors $Q_1=$ (0, 0, 1.5), $Q_2=$ (1, 0, 0.5), and $Q_3=$ (2, 0, 0.5) for $T=$ 1.5 and 300 K with $\theta_1=90\degree$, $\theta_2=27.6\degree$, and $\theta_3=14.6\degree$, respectively [Figs. 5(a)-(f)]. Here, $M_z$ always corresponds to the in-plane component $M_{\perp}$ for all three geometries, whereas the composition of $M_y$ varies with $Q$ as: $M_y(Q_1)=M_{\parallel}$, $M_y(Q_2)=0.21M_{\parallel} + 0.79M_{c}$, and $M_y(Q_3)=0.06M_{\parallel} + 0.94M_{c}$ \cite{supplementary}. From the energy scans at $Q_1$, it is evident that the spin excitations are isotropic within the $ab$ plane (i.e., the Fe kagome plane) as $M_y$ and $M_z$ are essentially superposed [Figs. 5(a) and 5(d)]. We therefore use $M_{ab}$ to denote the two equivalent in-plane components $M_{\perp}$ and $M_{\parallel}$. In the energy scans at $Q_3$, $M_y$ almost represents $M_c$. This component is nearly zero across most energies at 1.5 K, apart from weak intensity near 3 meV [Figs. 5(c) and 5(g)]. At 300 K, $M_y$ becomes nonzero below $E=$ 14 meV, yet remains far smaller than $M_z$ [Fig. 5(f)]. A similarly strong anisotropy between $M_{ab}$ and $M_{c}$ is observed in the energy scans at $Q_2$, although both components are weaker there [Figs. 5(b) and (e)]. 

To corroborate this spin anisotropy, we also performed $Q$ scans around $Q_3$ at $E=$ 3, 10, and 20 meV for $T=$ 1.5 and 300 K [Figs. 5(g)-5(l)]. Indeed, $M_y$ is consistent with zero within error bars at $E=$ 10 and 20 meV for $T=$ 1.5 K, while $M_z$ is well described by Gaussian fits. Notably, the low-energy portion ($E<$ 10 meV) of $M_z$ is strongly enhanced at 300 K but drops abruptly near 2 meV [Figs. 5(d) and 5(f)], while its high-energy part disperses into two peaks at 20 meV [Figs. 5(i) and 5(l)], in agreement with the unpolarized results. The temperature evolution of $M_y$ and $M_z$ at $Q_3$ for $E=$ 3, 10 and 20 meV provides further insight into the spin anisotropy  (Fig. 6): while $M_z$ at $E=$ 3 and 10 meV rises rapidly upon warming from 1.5 to 300 K, $M_y$ increases only gradually at these energies [Figs. 6(a)-(d)]. At $E=$ 20 meV, $M_z$ is weakly temperature dependent, and $M_y$ remains vanishingly small, close to zero [Figs. 6(e) and 6(f)].  

\begin{figure}[tbp]
	\includegraphics[width=0.45\textwidth]{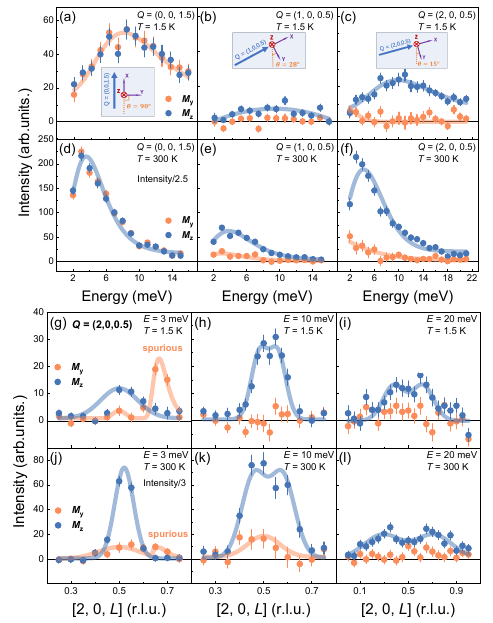}
	\caption{Polarized INS results collected at IN20.
		(a)-(f) Energy dependence of $M_y$ and $M_z$ at $Q$ = (0, 0, 1.5), (1, 0, 0.5) and (2, 0, 0.5) for $T$ = 1.5 and 300 K. Insets: Schematic crystallographic $Q$ directions and the spin-polarization directions ($x$, $y$, $z$) in reciprocal space. All solid lines are guides to the eye. 
		(g)-(l) $Q$ dependence of $M_y$ and $M_z$ along [2, 0, $L$] at $E$ = 3, 10, and 20 meV for $T$ = 1.5 and 300 K. The solid lines are Gaussian peak fittings. Spurious peaks are also observed for the case of $E=$ 3 meV.
	}
\end{figure}

\begin{figure}[tbp]
	\includegraphics[width=0.42\textwidth]{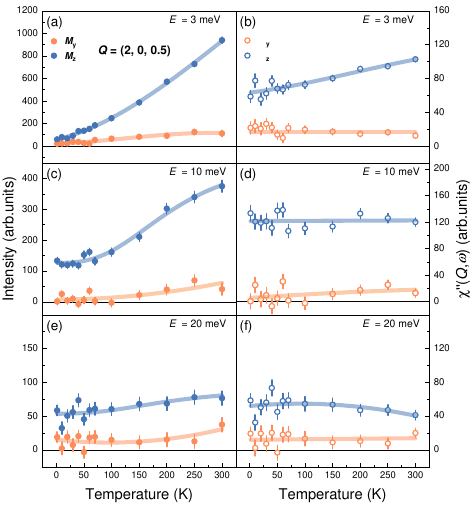}
	\caption{Polarized INS results at $Q$ = (2, 0, 0.5) collected at IN20.
		(a), (c), (e) Temperature dependence of $M_y$ and $M_z$ at $E=$ 3, 10, and 20 meV, and 
		(b), (d), (f) corresponding $\chi^{\prime\prime}_y$ and $\chi^{\prime\prime}_z$ with Bose factor corrections. All solid lines are guides to the eye. 
	}
\end{figure}

\section{Discussion}

Finally, we discuss the possible origin of the UAHE in Y$_{0.5}$Fe$_3$Sn$_3$. Previous study on YbFe$_6$Ge$_6$ suggested that the Yb-Fe magnetic coupling induces a spin reorientation from $c$-axis alignment to $ab$-plane arrangement, and that this reorientation is crucial for the UAHE emerging below $T_{SR} << T_N$. The reorientation renders the spins far more susceptible to fluctuations within the $ab$ plane, thereby generating gapless spin excitations that produce a nonzero average SSC under an in-plane magnetic field. In the present case of Y$_{0.5}$Fe$_3$Sn$_3$, however, Y atoms are nonmagnetic and no spin reorientation occurs at zero field (Fig. 1). We therefore argue that a spontaneous reorientation is unnecessary for the UAHE. It is important to recognize that both YbFe$_6$Ge$_6$ and Y$_{0.5}$Fe$_3$Sn$_3$ preserve the $\mathcal{IT}$ symmetry in their collinear AF states. This symmetry rules out a static nonzero SSC at zero field and ensures that any local SSC from spin fluctuations cancels out due to oppositely signed contributions on symmetry-related sites.
	
Under an in-plane magnetic field $\mathbf{B}\parallel ab$, however, the $\mathcal{IT}$ symmetry is lifted, and a fundamentally different scenario emerges. In the collinear $c$-axis AF state of Y$_{0.5}$Fe$_3$Sn$_3$, the spin excitations at low temperature are dominated by the gapless in-plane component $M_{ab}$ (Figs. 5 and 6), indicating that the Fe spins can be tilted into the kagome plane by the magnetic field at a small energy cost. The system thus develops a canted noncollinear spin configuration either in the static order or in dynamic fluctuations, in which the Fe moments acquire a finite in-plane component while retaining their dominant $c$-axis antiparallel arrangement. Crucially, the three Fe sublattices on the kagome net possess distinct local environments, such that their canting directions and magnitudes are inequivalent. It breaks the local chiral symmetry and forms a nonzero static SSC or time-averaged SSC, which in turn acts as an emergent magnetic field in real space and generates an additional Hall response beyond the conventional AHE. We note that the gapless in-plane excitations coexist with a static $c$-axis easy axis in this system, a situation that can arise when the intralayer Fe-Fe exchange is nearly $XY$-like within the kagome plane while a weak interlayer coupling or single-ion anisotropy pins the ordered moments along the $c$ axis. The strong spin anisotropy provides a natural explanation for why the UAHE is observed for $\mathbf{B}\parallel ab$ and not for $\mathbf{B}\parallel c$. A $c$-axis field preserves the collinear AF configuration and keeps the in-plane spin components quenched. The $\mathcal{IT}$ symmetry thus remains effectively unbroken with respect to the kagome-plane chirality, always yielding zero net SSC and no UAHE.

The temperature dependence of the UAHE, which disappears above 150 K, can be understood within the same framework. As shown by our polarized neutron data, the spin excitations become increasingly isotropic upon warming, but the in-plane component $M_{ab}$ gradually opens a gap at low energies (Figs. 5 and 6), which suppresses the in-plane dynamical susceptibility and thereby reduces the field-induced canting angle. The concomitant thermal population of $M_{c}$ does not compensate for this suppression, as $c$-axis fluctuations are orthogonal to the required tilting direction. Consequently, the net static/dynamic SSC and the associated UAHE are suppressed at elevated temperatures, vanishing near the threshold identified in our transport measurements.

Y$_{0.5}$Fe$_3$Sn$_3$ therefore constitutes a distinctive example in which a collinear $c$-axis AF order with highly anisotropic spin excitations supports a chirality-driven AHE under an in-plane magnetic field, without requiring either a spontaneous spin reorientation or the presence of magnetic rare-earth ions. This sublattice-canting mechanism broadens the known routes to unconventional Hall responses in kagome metals, extending beyond the established paradigms of momentum-space Berry curvature, noncollinear magnetism, and skyrmion-driven topological Hall effects.

\section{Conclusions}

In conclusion, we have performed neutron scattering measurements on the collinear $c$-axis AF kagome metal Y$_{0.5}$Fe$_3$Sn$_3$ ($T_N =$ 551 K), where an unconventional AHE emerges under $\mathbf{B}\parallel ab$. Elastic neutron scattering confirms that Fe moments retain $c$-axis alignment down to 1.5 K without spin reorientation. Inelastic scattering reveals a Y-shape spin-wave dispersion with a two-gap structure below 10 meV. Polarized measurements show strongly anisotropic excitations: gapless in-plane $M_{ab}$ dominates at 1.5 K, while $M_{c}$ is nearly absent and gradually emerges upon warming. We propose that the in-plane field induces sublattice-inequivalent canting of Fe moments either in the static order or dynamic fluctuations, producing a net scalar spin chirality that drives the UAHE. The suppression of UAHE above 150 K is attributed to the low-energy gap opening in $M_{ab}$, which reduces the field-induced canting. Y$_{0.5}$Fe$_3$Sn$_3$ thus exemplifies chirality-driven AHE in a collinear AF host without spontaneous spin reorientation.

\begin{acknowledgments}
This work is supported by the National Key Research and Development Program of China (Grants No.~2022YFA1403800, No.~2023YFA1406100, and No.~2025YFE0202100) and the National Natural Science Foundation of China (Grants No.~12574165, No.~12274444 and No.~12274438). F.P. and Y.Z. acknowledge the GNeuS postdoctoral funding from the European Union's Horizon 2020 Research and Innovation Programme under Marie Skłodowska-Curie Grant agreement No.~101034266. This work is based on experiments performed at the Swiss Spallation Neutron Source SINQ, Paul Scherrer Institute, Villigen, Switzerland (Proposal 20230521) and the Institut Laue-Langevin, Grenoble, France (Proposal 4-01-1854).

\end{acknowledgments}

\section*{Data availability}
The data collected from IN20 are available online \cite{IN20Exp2025}. Other data are available from the authors upon reasonable request.

\bibliographystyle{apsrev4}

\newpage
\appendix

\begin{center}
	{\bf Supplemental Materials}
\end{center}

\begin{center}
	{\bf A. Analyses on Hall Resistivity}
\end{center}

The Hall resistivity $\rho_{yx}$ has three contributions:  $\rho_{yx}=\rho^O_{yx}+\rho^A_{yx}+\Delta\rho^A_{yx}$, where $\rho^O_{yx}$ is the ordinary Hall resistivity linear in the applied magnetic field $B$, $\rho^A_{yx}$ is the conventional anomalous Hall resistivity arising from the net magnetization ($\rho^A_{yx}\propto M$), and $\Delta\rho^A_{yx}$ represents the unconventional anomalous Hall effect not captured by the first two terms \cite{Yao2025,wwang2022,Smejkal2022}. With the magnetization data in Fig. 2(e), we have fitted $\rho_{yx}$ by the combined term $aB+bM(B)$, where $a$ is negative due to the single-band electron-type charge carrier \cite{Liu2023}. The unconventional anomalous Hall resistivity is obtained from the difference between the raw data and the fitting curve: $\Delta\rho^A_{yx}=\rho_{yx}-(aB+bM)$ (Fig. S1). It is clear for a nonzero $\Delta\rho^A_{yx}$ below 2 T at low temperatures. The unconventional anomalous Hall conductivity can be deduced from $\Delta\sigma^A_{yx}=\Delta\rho^A_{yx}/(\rho^2_{xx}+\rho^2_{yx})$ from the data in Fig. 2 and Fig. S1.

\begin{center}
	{\bf B. Unpolarized neutron scattering experiments}
\end{center}

Unpolarized neutron scattering experiments were performed on the thermal triple-axis spectrometer EIGER at the Paul Scherrer Institut, Switzerland \cite{UStuhr2017}. The final neutron energy was fixed at $E_f=14.7$ meV ($k_{\mathrm{f}}=2.662$ \AA$^{-1}$) in energy-loss mode, with a pyrolytic graphite filter and a double-focusing monochromator. The scattering plane was $[H, 0, 0]\times[0, 0, L]$. The hexagonal unit cell with $a=b= 5.373$ \AA, $c=4.453$ \AA, $\alpha=\beta=90^{\circ}$, and $\gamma=120^{\circ}$ was used to define $\mathbf{Q}=H\mathbf{a}^*+K\mathbf{b}^*+L\mathbf{c}^*$, where $(H,K,L)$ are in reciprocal lattice units and $a^*=b^*=4\pi/\sqrt{3}a$, $c^*=2\pi/c$. About 4.6 g of single crystals were co-aligned with hydrogen-free glue on aluminum plates. Rocking scans of the $(3, 0, 0)$ and $(0, 0, 2)$ nuclear Bragg peaks show full widths at half maximum below $2^{\circ}$ for both in-plane and out-of-plane orientations (Fig. S2). In addition to the elastic scans along $[2, 0, L]$ discussed in the main text, scans along $[1, 0, L]$ show the same $c$-axis AF trend, with intensities at $Q=(1, 0, \pm0.5)$ decreasing upon warming. The integrated intensities of the elastic scans are summarized in Fig. 3(g) of the main text.

The spin excitation spectra were extracted from the constant-energy scans ($Q$-scans)shown in Fig. S3. The scattering cross section $S(\mathbf{Q}, E)$, obtained from the raw intensities by subtracting a linear background, was converted to the imaginary part of the dynamic susceptibility, $\chi^{\prime \prime}(\mathbf{Q}, E)$, after dividing by the Bose population factor:
\begin{equation}
	\chi^{\prime \prime}(\mathbf{Q}, E) = \left[1 - \exp(-E/k_B T)\right] S(\mathbf{Q}, \omega).
\end{equation}
For scans along $[H, 0, 0.5]$ [Figs. S3(a), S3(b)] and $[2, 0, L]$ [Figs. S3(c), S3(d)] at $T=1.5$ and 300 K, $\chi^{\prime\prime}(\mathbf{Q},E)$ was fitted with Gaussian peak functions on a zero background. A single Gaussian was sufficient for $E\leq10$ meV, whereas two Gaussians were used for $E\geq 14$ meV. The fitted peak centers define the spin-wave dispersions plotted in Figs. 3(a)-3(d) of the main text. The two-dimensional energy-momentum maps were generated by interpolation to display the excitation intensity.

Raw energy scans at $Q=(2, 0, 0.5)$ for $T=1.5$, 50, 100, 200, and 300 K are shown in Fig. S4(a)-S3(e). For comparison, an energy scan at $Q=(2, 0, 1.5)$ and $T=300$ K is shown in Fig. S4(f). Solid symbols denote the intensities at the magnetic peak centers, and open symbols denote backgrounds estimated from the average intensities at $(2, 0, 0)$ and $(2, 0, 1)$. After point-by-point background subtraction, the energy dependence of $S(\mathbf{Q},E)$ is obtained in Fig. S4(g) and converted to $\chi^{\prime\prime}(\mathbf{Q},E)$ in Fig. 4(e) of the main text. The temperature dependence at $E=3$ and 10 meV are shown in Fig. S4(h) and Fig. S4(i). Their deviation from the scaled Bose factors, $f/[1-\exp(-E/k_BT)]$, suggests combined contributions to the low-energy spin excitations.

\begin{center}
	{\bf C. Polarized neutron scattering analysis}
\end{center}

Polarized neutron scattering separates magnetic scattering from nuclear scattering because magnetic fluctuations or static moments perpendicular to $\mathbf{Q}$ can flip the neutron spin \cite{Lipscombe2010,NQureshi2012,HLuo2013,CWang2013,YSong2017,CLiu2022}. The polarized neutron scattering experiments were performed on the thermal triple-axis spectrometer IN20 at the Institut Laue-Langevin, France. Polarized neutrons were generated by a vertically focusing Heusler monochromator and analyzed by a vertically focusing Heusler analyzer, with a fixed final energy $E_f=14.7$ meV \cite{IN20Exp2025}. The CryoPAD system enabled spherical polarization analysis in a zero-field environment below 10 mG, suppressing the influence of the Earth's magnetic field \cite{Berna2005}.

The scattering plane was again $[H, 0, 0]\times[0, 0, L]$. The neutron polarization axes were defined as $x\parallel\mathbf{Q}$, $y\perp\mathbf{Q}$ within the scattering plane, and $z$ perpendicular to the scattering plane, as shown in Fig. 1(g) and Fig. 3(a)-3(c) of the main text. At selected energies and momentum transfers, we measured six cross sections: $\sigma_{xx}^{\mathrm{SF}}$, $\sigma_{yy}^{\mathrm{SF}}$, $\sigma_{zz}^{\mathrm{SF}}$, $\sigma_{xx}^{\mathrm{NSF}}$, $\sigma_{yy}^{\mathrm{NSF}}$, and $\sigma_{zz}^{\mathrm{NSF}}$. These correspond to spin-flip (SF) and non-spin-flip (NSF) processes for neutron polarizations along the $x$, $y$, and $z$ axes. In this $xx$, $yy$, and $zz$ polarization geometry, the cross sections are \cite{Lipscombe2010,NQureshi2012,HLuo2013,CWang2013,YSong2017,CLiu2022}:
\begin{equation}
	\left\{
	\begin{aligned}
		\sigma_{xx}^{\mathrm{SF}} &= \frac{R}{R+1} M_{y} + \frac{R}{R+1} M_{z} + \frac{2}{3} \sigma_{\mathrm{Inc}}  \\
		\sigma_{yy}^{\mathrm{SF}} &= \frac{1}{R+1} M_{y} + \frac{R}{R+1} M_{z} + \frac{2}{3} \sigma_{\mathrm{Inc}}    \\
		\sigma_{zz}^{\mathrm{SF}} &= \frac{R}{R+1} M_{y} + \frac{1}{R+1} M_{z} + \frac{2}{3} \sigma_{\mathrm{Inc}}
	\end{aligned}
	\right.
\end{equation}
and,
\begin{equation}
	\left\{
	\begin{aligned}
		\sigma_{xx}^{\mathrm{NSF}} &= \frac{1}{R+1} M_{y} + \frac{1}{R+1} M_{z}  + \frac{R}{R+1} \sigma_{\mathrm{Nuc}} + \frac{1}{3} \sigma_{\mathrm{Inc}} \\ 
		\sigma_{yy}^{\mathrm{NSF}} &= \frac{R}{R+1} M_{y} + \frac{1}{R+1} M_{z}  + \frac{R}{R+1} \sigma_{\mathrm{Nuc}} + \frac{1}{3} \sigma_{\mathrm{Inc}} \\ 
		\sigma_{zz}^{\mathrm{NSF}} &= \frac{1}{R+1} M_{y} + \frac{R}{R+1} M_{z}  + \frac{R}{R+1} \sigma_{\mathrm{Nuc}} + \frac{1}{3} \sigma_{\mathrm{Inc}}
	\end{aligned}
	\right.
\end{equation}
respectively.
Here, $M_y$ and $M_z$ are the magnetic scattering components along the $y$ and $z$ spin directions, $\sigma_{\mathrm{Nuc}}$ is the coherent nuclear scattering, and $\sigma_{\mathrm{Inc}}$ is the isotropic incoherent background. The flipping ratio, $R=\sigma_{\mathrm{Nuc}}^{\mathrm{NSF}}/\sigma_{\mathrm{Nuc}}^{\mathrm{SF}}$, measures the leakage of nuclear Bragg intensity into the SF channel and quantifies the polarization efficiency. In our experiment, $R\approx18$, so $R/(R+1)\approx1$ and $1/(R+1)\approx0$. Solving the above equations gives:
\begin{equation}
	\left\{
	\begin{aligned}
		M_y &= [(\sigma_{xx}^{\mathrm{SF}} - \sigma^{\mathrm{SF}}_{yy}) + (\sigma^{\mathrm{NSF}}_{yy} - \sigma^{\mathrm{NSF}}_{xx})]/2,\\
		M_z &= [(\sigma^{\mathrm{SF}}_{xx} - \sigma^{\mathrm{SF}}_{zz}) + (\sigma^{\mathrm{NSF}}_{zz} - \sigma^{\mathrm{NSF}}_{xx})]/2,\\
		\sigma_{\mathrm{Nuc}} &= (\sigma^{\mathrm{NSF}}_{xx} + \sigma^{\mathrm{NSF}}_{yy}+ \sigma^{\mathrm{NSF}}_{zz})/3 - (\sigma^{\mathrm{SF}}_{xx} + \sigma^{\mathrm{SF}}_{yy} + \sigma^{\mathrm{SF}}_{zz})/6,\\
		\sigma_{\mathrm{Inc}} &= (\sigma^{\mathrm{SF}}_{yy} + \sigma^{\mathrm{SF}}_{zz} - \sigma^{\mathrm{SF}}_{xx})\times 3/2.
	\end{aligned}
	\right.
\end{equation}
This analysis estimates $M_y$ and $M_z$ from both SF and NSF channels, reducing statistical uncertainty. In practice, the three SF channels alone can also determine $M_y$ and $M_z$ \cite{HLuo2013,CWang2013,YSong2017,CLiu2022}. The components $M_y$ and $M_z$ are related to the spin fluctuations along the crystallographic axes by:
\begin{equation}
	\left\{
	\begin{aligned}
		M_y(\mathbf{Q}) &= M_{\parallel}\sin^2(\theta) + M_c\cos^2(\theta), \\
		M_z &= M_{\perp}.
	\end{aligned}
	\right.
\end{equation}
$M_c$ is the $c$-axis component, while $M_{\parallel}$ and $M_{\perp}$ are the two in-plane components along $[H, 0, 0]$ ($\mathbf{a}^*$) and $[-0.5K, K, 0]$ ($\mathbf{Q}\perp\mathbf{a}^*$), respectively. The angle $\theta$ is defined between the measured momentum transfer $\mathbf{Q}$ and the crystalline $[0,0,L]$ direction ($\mathbf{c}^*$). Thus, $M_{\parallel}$, $M_{\perp}$, and $M_c$ can be resolved by measuring at two or more nonparallel wave vectors. Cross sections were measured at $Q_1=(0, 0, 1.5)$, $Q_2=(1, 0, 0.5)$, and $Q_3=(2, 0, 0.5)$, with $\theta_1=90\degree$, $\theta_2=27.6\degree$, and $\theta_3=14.6\degree$. Therefore \cite{HLuo2013,CWang2013,YSong2017,CLiu2022}:

\begin{equation}
	\left\{
	\begin{aligned}
		M_y(Q_1) &= \sin^2(90\degree)M_{\parallel} + \cos^2(90\degree)M_c\\ &=M_{\parallel}, \\
		M_y(Q_2) &= \sin^2(27.6\degree)M_{\parallel} + \cos^2(27.6\degree)M_c\\ &=0.21M_{\parallel} + 0.79M_{c}, \\
		M_y(Q_3) &= \sin^2(14.6\degree)M_{\parallel} + \cos^2(14.6\degree)M_c\\ &=0.06M_{\parallel} + 0.94M_{c}.
	\end{aligned}
	\right.
\end{equation}

For all three wave vectors, $M_z$ corresponds to $M_{\perp}$ because $M_{\perp}$ is perpendicular to the scattering plane. In practice, $M_z(Q_1)$, $M_z(Q_2)$, and $M_z(Q_3)$ may differ because of the magnetic form factor and instrumental resolution. These effects can be accounted for by temperature-independent scale factors when comparing intensities at different $\mathbf{Q}$ positions. At $Q_1=(0, 0, 1.5)$, $M_y$ and $M_z$ are equal within error bars [Figs. 5(a), 5(d) of the main text], indicating isotropic in-plane spin excitations with $M_{\parallel}=M_{\perp}$. We therefore denote both in-plane components by $M_{ab}$ in the main text. At $Q_2$ and especially at $Q_3$, $M_y$ is dominated by the $c$-axis component $M_c$, enabling a direct comparison between in-plane and out-of-plane spin excitations.

The raw polarized neutron data and the extracted $\sigma_{\mathrm{Nuc}}$ and $\sigma_{\mathrm{Inc}}$ are shown in Fig. S5-Fig. S9. Energy scans at $Q_1$, $Q_2$, and $Q_3$ were measured at $T=1.5$ K (Fig. S5) and 300 K (Fig. S6). The extracted $\sigma_{\mathrm{Nuc}}$ and $\sigma_{\mathrm{Inc}}$ are nearly zero for these energy scans, indicating weak phonon and incoherent backgrounds. Constant-energy $Q$ scans along $[2, 0, L]$ were measured at the same temperatures for selected energies, as shown in Fig. S7 and Fig. S8. At 1.5 K, an additional $[1,0,L]$ scan at $E=10$ meV gives a similar but weaker signal than the corresponding $[2, 0, L]$ scan. Weak nuclear scattering is observed at $E=10$ and 20 meV near large $\mathbf{Q}$ around $(2, 0, 0.5)$, while the weak extra peak in the $E=3$ meV scans is magnetic in origin. The temperature dependence at $Q=(2, 0, 0.5)$ and $E=3$, 10, and 20 meV is summarized in Fig. S0. The magnetic intensity increases on warming to 300 K, whereas $\sigma_{\mathrm{Inc}}$ remains nearly temperature independent.

\begin{center}
	{\bf D. Magnetically ordered moment}
\end{center}

The magnetically ordered moment is estimated by comparing the intensity of a magnetic Bragg peak with that of a nuclear Bragg peak measured under the same instrumental conditions. In this way, the nuclear reflection provides an absolute structural reference, while the magnetic reflection carries the information on the ordered moment of Fe. In elastic neutron scattering, the corresponding integrated intensities, $I_N$ and $I_M$, are written as \cite{Collins1967,hqluo2012,xylu2013,dgong2018,wlzhang2019}:
\begin{equation}
	I_N =
	A N_N \frac{(2\pi)^2}{V_N}
	\times
	\frac{|F_N(\bm{Q})|^2}{\sin(2\theta_N)} ,
\end{equation}
and
\begin{equation}
	I_M =
	A N_M \frac{(2\pi)^2}{V_M}
	\times
	\frac{|F_M(\bm{Q})|^2}{2\sin(2\theta_M)} .
\end{equation}

Here, $N_N$ and $N_M$ are the numbers of atoms in the nuclear and magnetic unit cells, and $V_N$ and $V_M$ are the corresponding unit-cell volumes. For Y$_{0.5}$Fe$_3$Sn$_3$, the magnetic unit cell is doubled along the magnetic propagation direction, so that $N_M=N_N/2=4$ and $V_M=2V_N$. The structure factors $F_N(\bm{Q})$ and $F_M(\bm{Q})$ describe how strongly a given nuclear or magnetic Bragg reflection contributes at momentum transfer $\bm{Q}$, while $2\theta_N$ and $2\theta_M$ account for the corresponding scattering geometry through the Lorentz factor.

The magnetic structure factor is written as
\begin{equation}
	F_M(\bm{Q}) =
	p S \sum_i (-1)^i e^{\mathrm{i}\bm{Q}\cdot \bm{d}_i},
\end{equation}
where
\begin{equation}
	p =
	0.2659 \times 10^{-12}\ \mathrm{cm} \times g \times f_M(\bm{Q}) .
\end{equation}
% Here, $f_M(\bm{Q})$ is the magnetic form factor of Fe$^{2+}$, parameterized as
The quantity $f_M(\bm{Q})$ is the magnetic form factor of Fe$^{2+}$, which describes the reduction of magnetic scattering with increasing $|\bm{Q}|$ due to the spatial extent of the Fe magnetic electrons. It is parameterized as
\begin{equation}
	f_M(\bm{Q}) =
	A e^{-a Q^2/16\pi^2}
	+
	B e^{-b Q^2/16\pi^2}
	+
	C e^{-c Q^2/16\pi^2}
	+
	D .
\end{equation}

In these expressions, $S$ is the ordered moment per Fe ion contributing to the magnetic Bragg scattering at $\bm{Q}$, and $\bm{d}_i$ gives the position of the $i$th magnetic ion in the magnetic unit cell. The phase factor $e^{\mathrm{i}\bm{Q}\cdot \bm{d}_i}$ determines how the moments on different sites interfere at the selected reflection. The Land\'{e} factor is taken as $g=2$.

In our analysis, the contribution from two equivalent magnetic twin domains is included in Eq.~(8) through the extra factor of 2 in the denominator. Physically, this means that the measured magnetic intensity is shared by two symmetry-equivalent domain populations. After this domain effect and the structural and geometrical factors are taken into account, the ordered moment can be obtained by comparing the integrated intensities of the nuclear and magnetic Bragg peaks \cite{hqluo2012,xylu2013,dgong2018,wlzhang2019,Cadogan2000}:
\begin{equation}
	S =
	0.067
	\sqrt{
		\frac{I_M \sin 2\theta_M}
		{I_N \sin 2\theta_N}
	}
	\frac{|F_N|}{|f_M|}.
\end{equation}
Experimentally, we use two magnetic reflections, $(1, 0, 0.5)$ and $(2 ,0, 0.5)$, together with two nuclear reflections, $(1, 0, 1)$ and $(2, 0, 1)$, to evaluate the ordered moment. Each combination provides an independent estimate of $S$ through Eq.~(12). To reduces the uncertainty in a single reflection combination, the value quoted in the main text is the average of these two estimates.

\newpage

\begin{figure*}
	\renewcommand\thefigure{S1}
	\includegraphics[width=0.95\textwidth]{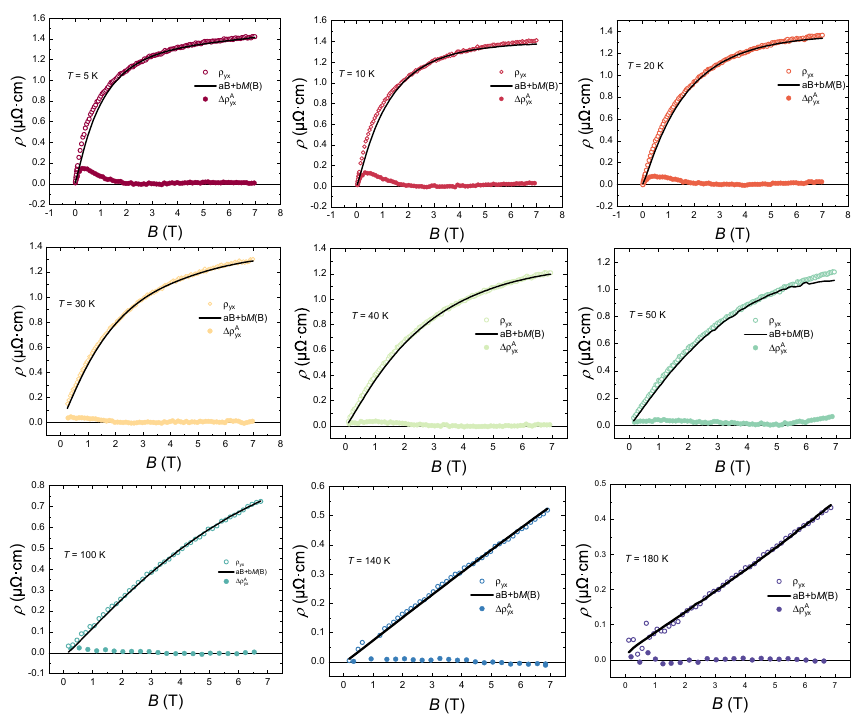}
	\caption{Fitting result of the Hall resistivity: $\rho_{yx}=\rho^O_{yx}+\rho^A_{yx}+\Delta\rho^A_{yx}$. Here the ordinary Hall resistivity $\rho^O_{yx}=aB$ , the conventional anomalous Hall resistivity $\rho^A_{yx}=bM(B)$, and the unconventional anomalous Hall resistivity $\Delta\rho^A_{yx}=\rho_{yx}-(aB+bM)$.
	}
	\label{S1}
\end{figure*}

\begin{figure*}
	\renewcommand\thefigure{S2}
	\includegraphics[scale = 1]{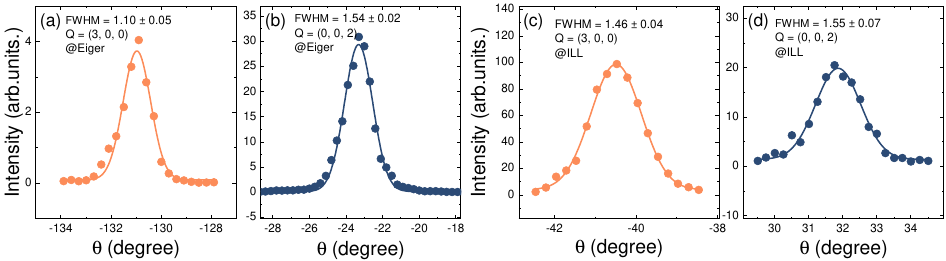}
	\caption{Rocking scans of the nuclear Bragg peaks used to characterize the crystal co-alignment. Panels (a) and (b) show scans through $Q=(3,0,0)$ and $(0,0,2)$ measured on EIGER, and panels (c) and (d) show the corresponding scans measured on IN20. Solid lines are Gaussian fits with constant backgrounds; the fitted FWHM is indicated in each panel.
	}
	\label{S2}
\end{figure*}

\begin{figure*}
	\renewcommand\thefigure{S3}
	\includegraphics[scale = 1.1]{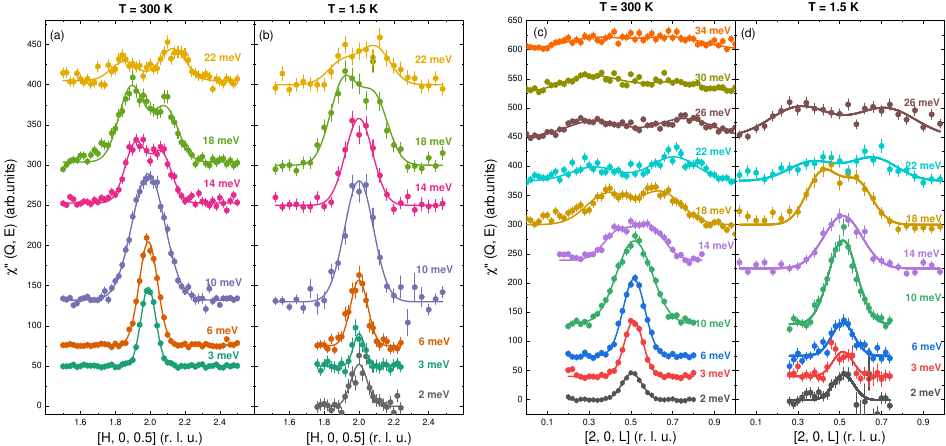}
	\caption{Q-scans along $[H,0,0.5]$ (left panels) and $[2,0,L]$ (right panels) at 300 and 1.5 K. Solid lines are Gaussian fits used to determine the peak positions of the Y-shaped spin-wave dispersions.
	}
\end{figure*}

\begin{figure*}
	\renewcommand\thefigure{S4}
	\includegraphics[scale = 1.2]{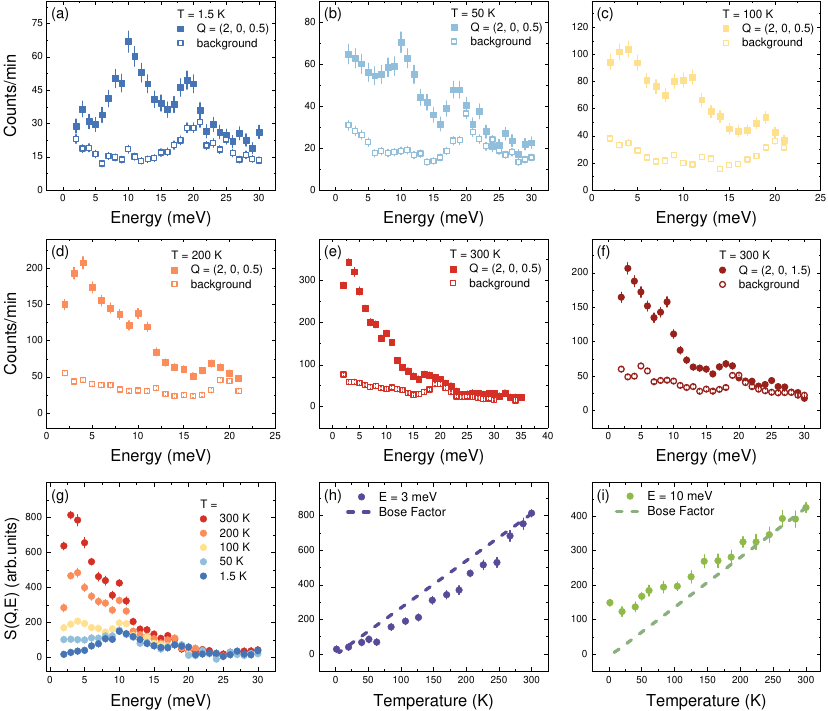}
	\caption{Unpolarized energy scans. (a)-(e) Energy scans at $Q=(2,0,0.5)$ from 1.5 to 300 K. (f) Energy scan at $Q=(2,0,1.5)$ and 300 K. Solid symbols are intensities at the magnetic peak center, and open symbols are backgrounds estimated from the average intensities at $(2,0,0)$ and $(2,0,1)$. (g) Background-subtracted magnetic scattering $S(\mathbf{Q},E)$ at $Q=(2,0,0.5)$. (h),(i) Temperature dependences at $E=3$ and 10 meV. Dashed lines are scaled Bose factors $f/[1-\exp(-E/k_BT)]$.
	}
\end{figure*}

\begin{figure*}
	\renewcommand\thefigure{S5}
	\includegraphics[scale = 1.2]{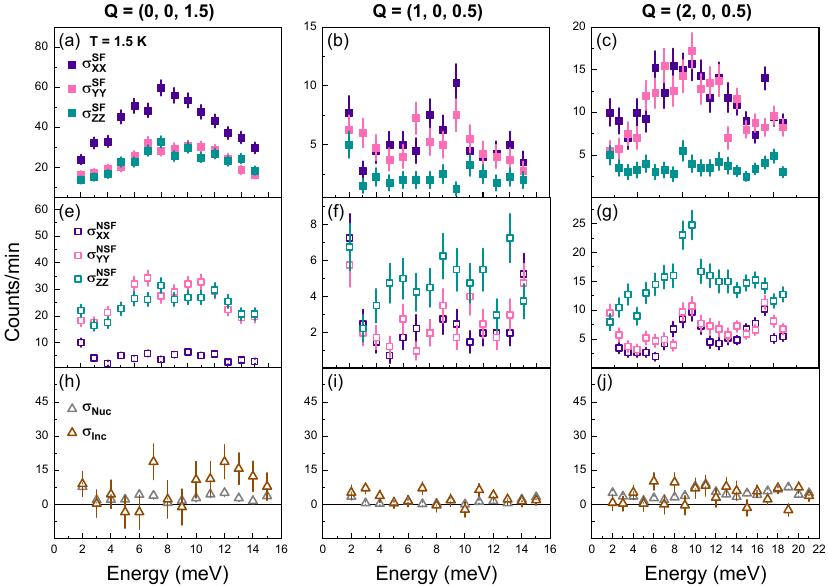}
	\caption{Polarized energy scans at $Q_1=(0,0,1.5)$, $Q_2=(1,0,0.5)$, and $Q_3=(2,0,0.5)$ at 1.5 K. (a)-(c) Raw SF cross sections $\sigma_{xx}^{\mathrm{SF}}$, $\sigma_{yy}^{\mathrm{SF}}$, and $\sigma_{zz}^{\mathrm{SF}}$. (d)-(f) Raw NSF cross sections $\sigma_{xx}^{\mathrm{NSF}}$, $\sigma_{yy}^{\mathrm{NSF}}$, and $\sigma_{zz}^{\mathrm{NSF}}$. (g)-(i) Extracted nuclear scattering $\sigma_{\mathrm{Nuc}}$ and incoherent background $\sigma_{\mathrm{Inc}}$.
	}
\end{figure*}

\begin{figure*}
	\renewcommand\thefigure{S6}
	\includegraphics[scale = 1.2]{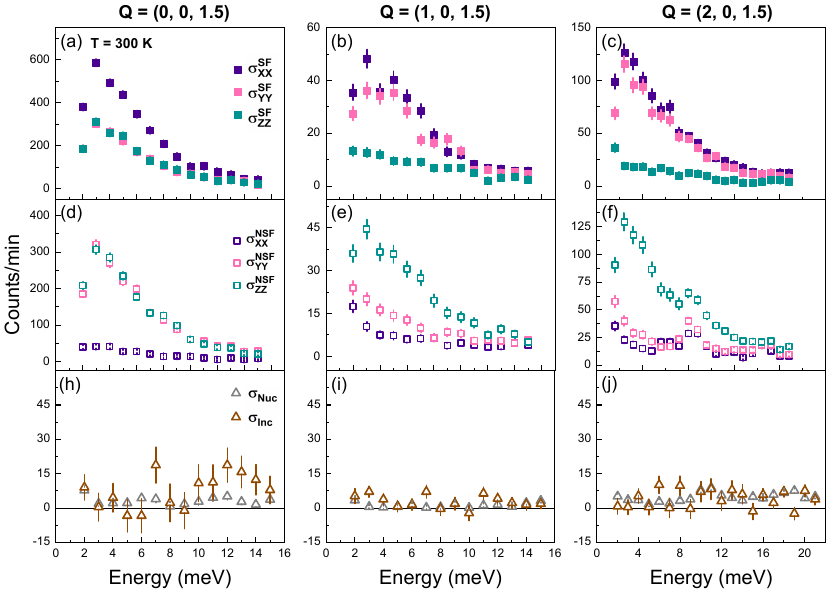}
	\caption{Polarized energy scans at $Q_1=(0,0,1.5)$, $Q_2=(1,0,0.5)$, and $Q_3=(2,0,0.5)$ at 300 K. (a)-(c) Raw SF cross sections. (d)-(f) Raw NSF cross sections. (g)-(i) Extracted $\sigma_{\mathrm{Nuc}}$ and $\sigma_{\mathrm{Inc}}$.
	}
\end{figure*}

\begin{figure*}
	\renewcommand\thefigure{S7}
	\includegraphics[scale = 1.1]{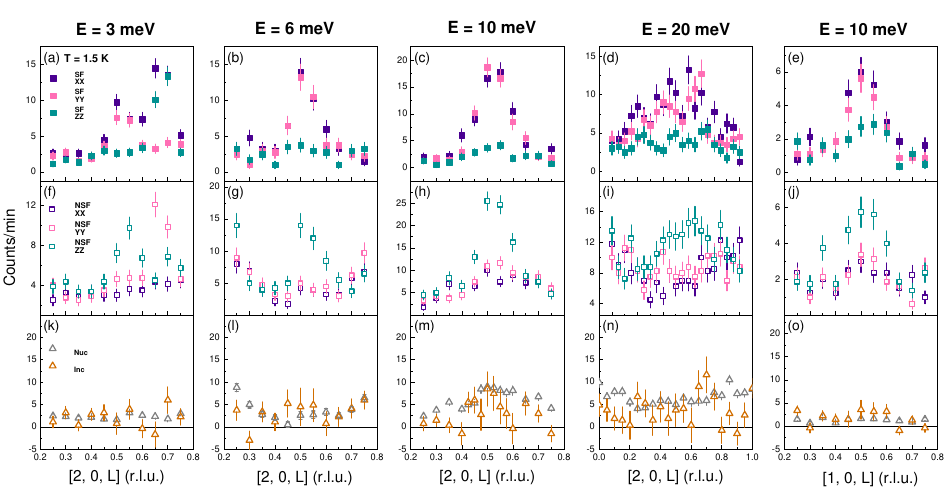}
	\caption{Polarized $Q$-scans at selected energies for $T=1.5$ K. The first four columns show scans along $[2,0,L]$ at $E=3$, 6, 10, and 20 meV, and the last column shows a scan along $[1,0,L]$ at $E=10$ meV. (a)-(e) Raw SF cross sections. (f)-(j) Raw NSF cross sections. (k)-(o) Extracted $\sigma_{\mathrm{Nuc}}$ and $\sigma_{\mathrm{Inc}}$.
	}
\end{figure*}

\begin{figure*}
	\renewcommand\thefigure{S8}
	\includegraphics[scale = 1.2]{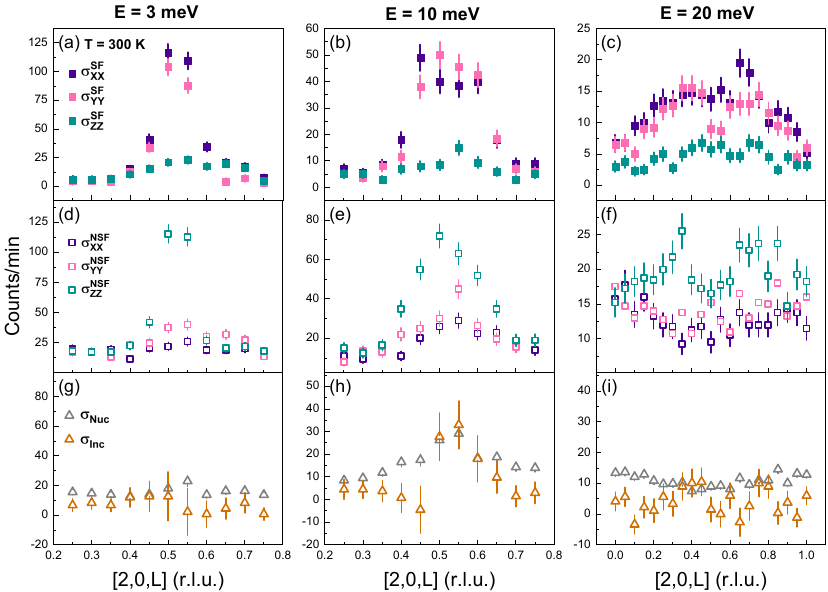}
	\caption{Polarized $Q$-scans along $[2,0,L]$ at $E=3$, 10, and 20 meV for $T=300$ K. (a)-(c) Raw SF cross sections. (d)-(f) Raw NSF cross sections. (g)-(i) Extracted $\sigma_{\mathrm{Nuc}}$ and $\sigma_{\mathrm{Inc}}$.
	}
\end{figure*}

\begin{figure*}
	\renewcommand\thefigure{S9}
	\includegraphics[scale = 1.2]{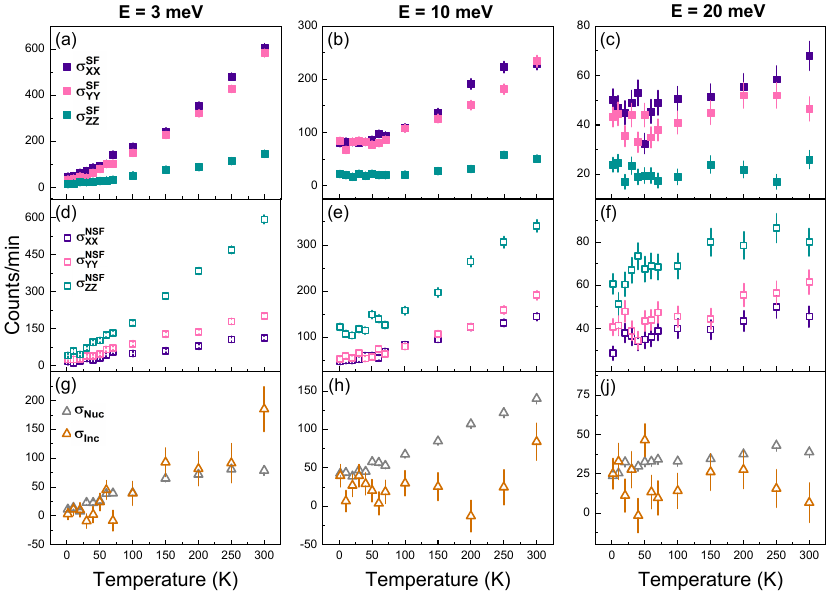}
	\caption{Temperature dependence of polarized neutron intensities at $Q=(2,0,0.5)$ and $E=3$, 10, and 20 meV. (a)-(c) Raw SF cross sections. (d)-(f) Raw NSF cross sections. (g)-(i) Extracted $\sigma_{\mathrm{Nuc}}$ and $\sigma_{\mathrm{Inc}}$.
	}
\end{figure*}

\end{document}